\documentclass[aps,prb,floatfix]{revtex4}
\usepackage{graphicx}

\begin{document}

\title{ First principles study of Sc, Ti and V doped Na$_n$(n =4, 5,
    6) clusters: Enhanced magnetic moments}

\author{Kalpataru Pradhan and Prasenjit Sen
\footnote{e-mail:prasen@hri.res.in}}
\affiliation{Harish-Chandra Research Institute, Chhatnag Road, Jhunsi, 
Allahabad 211019, INDIA.}

\author{J.\ U.\ Reveles and S.\ N.\ Khanna}
\affiliation{Department of Physics, Virginia Commonwealth University, Richmond VA, USA.}

\begin{abstract}
Theoretical studies on the geometry, electronic structure and spin
multiplicity of Sc, Ti and V doped Na$_n$ (n = 4, 5, 6) clusters have been
carried out within a gradient corrected density functional approach. Two
complementary approaches including all-electron calculations on free clusters,
and supercell calculations using planewave pseudopotential and projector
augmented wave formalisms have been carried out. It is shown that spin
magnetic moments of the transition metal atoms, the magnitude of host
polarization, and the sign of the host polarization all change with the number
of alkali atoms. In particular the transition metal atoms are shown to attain
spin moments that are higher than their atomic values. The role of
hybridization between the transition atom $d$-states and the alkali $sp$-states is
highlighted to account for the evolutions in the spin moments and host
polarization.

\end{abstract}

\maketitle

\section{Introduction}
Extensive research over the past two decades has shown that atomic clusters
consisting of two to a few hundred atoms constitute a new phase of matter with
distinct properties different from those of individual atoms on the one hand
and bulk matter on the other~\cite{01,02,03,04,05,06}. Of all the properties, the magnetic
properties in reduced sizes are probably the most fascinating~\cite{07}. Small clusters
of itinerant ferromagnetic solids Fe, Co, and Ni are found to possess enhanced
magnetic moments that can be around 30$\%$ higher than the bulk moments~\cite{08}. 
These clusters also exhibit super paramagnetic behavior as a consequence of
reduced magnetic anisotropy as opposed to the ferromagnetic behavior observed
in solids~\cite{08,09,10}. Small clusters of non-magnetic solids like Rh are found to be
magnetic with appreciable moments per atom that disappear as the size
approaches around 100 atoms~\cite{11,12}. Transition metal (TM) atoms doped in
clusters of other metals also show very interesting magnetic
properties. Recently, Li {\it et al. }~\cite{13} studied $3d$ TM doped Au$_6$ clusters both
experimentally, and theoretically through density functional theory (DFT)
calculations. In these clusters, the TM atoms were found to retain their
atomic moments. Later, Torres {\it et al. }~\cite{14} reported DFT calculations of TM doped
small cationic Au clusters in which the magnetic moment showed pronounced
odd-even effects as a function of the cluster size, and resulted in values
very sensitive to the geometrical environment. In all these cases, the moment
per atom in the clusters are less than or equal to the possible spin magnetic
moment in the individual atom. It has however been recently reported that
atomic and even magnetic moments beyond the atomic value on transition metal
atoms can be stabilized by depositing them on alkali metal surfaces and in the
interior of alkali hosts. For example, Bergmann and Song~\cite{15,16} have carried out
anomalous Hall effect measurements on systems consisting of Vanadium
impurities on Na and K thin films. Vanadium is non-magnetic in the solid
phase. Yet, the measurements by Bergman and Song indicate a large moment in
the range of 6.6 to 7 $\mu_B$ per atom. Sahu and Kleinman~\cite{17} have carried out first
principles theoretical studies on Vanadium inside alkali hosts using a
supercell approach. The theoretical studies suggest that the absolute value of
the moment may be less than that predicted by Bergman and Song. Both studies,
however, do confirm large spin moments on V sites indicating that alkali hosts
can stabilize and enhance spin moments on transition metal sites.

In this paper, we present an intriguing finding where the spin magnetic moment
of transition metal atom and the polarizability of the host atoms can be
controlled by changing the size of the alkali host. Our investigations cover
Sc, Ti, and V atoms deposited on alkali templates consisting of four, five and
six sodium atoms. The investigations employ first principles theoretical
studies within a gradient corrected density functional to incorporate the
exchange and correlation effects. We demonstrate that as the cluster size is
changed, the three impurities exhibit differing behavior. While the spin
moment of Sc and V atoms oscillate as the number of alkali atoms is changed,
the moment of Ti atom monotonically increases as the size is increased. A
detailed analysis of the atomic spin charges on Na atoms shows that while the host
polarization enhances the spin moment in some cases, it opposes that in
others. The ability to control the spin moment and the host polarization by
changing the metal atom and the size of the alkali cluster offers unique
opportunities for tuning magnetic moment and coupling in such systems.

\section{Computational Methods}

All the theoretical studies were carried out within the gradient corrected DFT
framework.  We carried out all-electron calculations with atom-centered
Gaussian type orbitals using the deMon2K code~\cite{18}. These calculations employed
the PW86 generalized gradient approximation (GGA) functional~\cite{19}, and the DZVP
basis sets optimized for gradient corrected exchange-correlation
functionals~\cite{20}. In order to avoid the calculation of four-center electron
repulsion integrals, the variational fitting of the Coulomb
potential~\cite{21,22} was
employed. The auxiliary density was expanded in primitive Hermite Gaussian
functions using the A2 auxiliary function set~\cite{23}. The exchange-correlation
potential was calculated with the orbital density. Structures were optimized
using the quasi-Newton method in internal redundant coordinates~\cite{24} for
different fixed spin multiplicities, and without any symmetry
constraints. Supplementary calculations were performed in order to eliminate
any uncertainty due to the basis set or the numerical procedure. We also
carried out plane-wave pseudopotential calculations using a supercell
approach. Here, an energy cutoff of 500 eV was used, and the cluster was
placed in a large cubic box of sides 20 \AA~ in order to reduce its interaction
with its images. The potential between the ion cores and the valence electrons
were expressed in terms of the Vanderbilt-type ultra-soft pseudopotential
(USPP)~\cite{25}. The exchange-correlation effects were treated with the PW91~\cite{26} GGA
functional. Brillouin zone integrations were carried out using only the
$\Gamma$-point. Structures were relaxed using the conjugate gradient (CG) method for
different fixed spin multiplicities, and without any symmetry constraints. We
re-calculate total energies of all clusters using projector augmented wave
(PAW) method in the same geometry as obtained from USPP calculations. These
calculations were carried out using VASP code~\cite{27,28,29}. It turned out that the
all-electron, USPP and PAW calculations all agreed on the lowest energy spin
multiplicity for each cluster. We found some differences in the energy
orderings and spin multiplicities of excited states between the plane-wave and
Gaussian type methods; however these differences are small in most of the
cases. In what follows we report the geometries and energy values based on
all-electron deMon2K calculations, unless otherwise mentioned.

\section{Results and Discussion}
\subsection{Pure Na clusters}
In order to benchmark our calculations, we first calculated structures and
binding energies (BE) of pure Na$_n$ clusters in the size range $n = 3,4,\ldots,10$ 
where previous calculations exist. Binding energy of a cluster is defined
as the energy gain in assembling the cluster from the isolated atoms. For pure
Na clusters this is given by the equation:
\begin{equation}
  \label{eq:ec}
  BE({\rm Na}_n) = nE({\rm Na}) - E({\rm Na}_n), 
\end{equation}
\noindent
where E(Na$_n$) represents the total energy of a Na$_n$ cluster, and E(Na) is the
total energy of an isolated Na atom. Note that according to this definition,
BE is a positive number for a bound structure, and a larger BE implies a more
stable structure.

In both all-electron and USPP methods we were able to reproduce the previously
reported structures of pure Na clusters~\cite{30}. In
particular, we reproduced the C$_{2v}$ structure for a pure Na$_5$ cluster, and for a
Na$_6$ cluster we found two close lying isomers with C$_{5v}$ and D$_{3h}$ symmetries. We
found the 3-dimensional (3D) C$_{5v}$ structure to be marginally ($\sim 0.03$ eV) lower
in energy than the planar D$_{3h}$ structure. Earlier calculations~\cite{30,31} also found
these two structures to be very close in energy with different methods giving
different energy ordering. We also found that the BE per Na atom shows a peak
at $n = 8$ relative to neighboring sizes. Enhanced stability of Na$_8$ with 8
valence electrons has been previously rationalized within a jellium model~\cite{32}
where the cluster corresponds to a magic species with filled electronic shell.

Having confirmed that our calculations reproduce the known structures and
energetics of pure Na clusters at small sizes correctly, we then studied TM
doping in Na$_n$ clusters for $n = 4-6$. At each size, we tried several initial
geometries and relaxed them for different fixed spin states without any
symmetry constraints. We now discuss the results of these calculations. Figure~\ref{fig:1}
presents the optimized geometries and Mulliken atomic spin charges of the ground state 
and the next higher lying isomer in energy for the series TMNa$_n$ ($n =$ 4, 5, and
6 and TM = Sc, Ti, and V), and in Table~\ref{Table:1}  we present the magnetic moment and
relative energies of these clusters.

\subsection{TMNa$_4$ clusters}
In the case of TMNa$_4$ clusters, we found two low lying structures for ScNa$_4$
which are shown in Figure~\ref{fig:1}. We found the same structure for the ground state
for ScNa$_4$, TiNa$_4$ and VNa$_4$ clusters. This structure is obtained by replacing
the Na atom with the highest coordination number in the C$_{2v}$ structure of a
pure Na$_5$ cluster by a TM atom. ScNa$_4$ and TiNa$_4$ clusters retain the atomic
moments of the Sc and Ti atoms, 1 and 2 $\mu_B$ respectively. What is intriguing
and amazing is that the VNa$_4$ cluster has an enhanced magnetic moment of 5 $\mu_B$,
while that of a V atom in the ground state is only 3 $\mu_B$. The microscopic
reason behind this enhanced magnetic moment is discussed later.

A summary of the results on these clusters are presented in Table~\ref{Table:1}. It is to
be noted, however, that both for TiNa$_4$ and VNa$_4$ clusters, the high and low
spin-multiplicity states are very close in energy, while the two structural
isomers of ScNa$_4$ are also close in energy.

\subsection{TMNa$_5$ clusters}
For TMNa$_5$ clusters, the low energy structures we found are: (i) a C$_{5v}$
structure with a TM atom incorporated at the center of a Na$_5$ pentagon for the
ground states of TiNa$_5$ and VNa$_5$; (ii) a square bipyramidal structure having
C$_{4v}$ symmetry for the ground state of ScNa$_5$ and the low lying isomer of VNa$_5$
and, (iii) a structure with C$_s$ symmetry that is obtained by distorting the C$_{4v}$
structure for the low lying isomers of ScNa$_5$ and TiNa$_5$. These structures are
shown in Figure~\ref{fig:1} and the main results of our calculations for TMNa$_5$ clusters
are presented in Table~\ref{Table:1}. 

Although the ScNa$_5$ cluster has no net moment in its lowest energy structure,
we found a nearly isoenergetic state with magnetic moment of 2 $\mu_B$ only 0.003
eV higher in energy. TiNa$_5$ and VNa$_5$ clusters presented large moments of 3 and
4 $\mu_B$ respectively in their ground states. The next higher energy
structures/multiplets of TiNa$_5$ and VNa$_5$ are separated by significant energy
differences 0.1 eV and 0.3 eV respectively from the ground states. As we
discuss below, similar large moments are found in Ti and V doped Na$_6$ clusters
and such large magnetic moments seem to be a feature of early TM doped small
Na clusters. 

\subsection{TMNa$_6$ clusters}
For TMNa$_6$ clusters, we studied both planar and 3D structures having various
symmetries within plane-wave USPP and PAW methods. In a planar geometry, we
placed the TM at the center of a Na$_6$ hexagon. In this geometry, ScNa$_6$ has a
moment of 3 $\mu_B$ , while the Ti and V doped clusters have moments as large as 4
$\mu_B$ and 5  $\mu_B$ respectively.

As for 3D structures for TMNa$_6$ clusters, we found stable structures with
C$_{5v}$, and C$_{3v}$ or C$_{s}$ symmetries as shown in Figure~\ref{fig:1}. In all the cases, the C$_{5v}$
structure turned out to be the ground state. This pentagonal bipyramid
structure is exactly what was found for the MgNa$_6$ cluster~\cite{33} and is also
equivalent to attaching a TM atom to the C$_{5v}$ structure for a pure Na$_6$
cluster. 

The 3D structures turned out to be significantly more favorable in energy than
the planar structure. The ground states of ScNa$_6$, TiNa$_6$ and VNa$_6$ present
moments of 1, 4 and 5 $\mu_B$ respectively. The next higher energy low lying
isomers of TiNa$_6$ and VNa$_6$ have magnetic moments of 2 and 3 $\mu_B$, which are
separated from the ground states by 0.03 eV and 0.13 eV respectively.  These
results are presented in Table~\ref{Table:1}. Li {\it et al.}~\cite{13} have claimed that Ti and V doped
Au$_6$ clusters have moments of 2 and 3 $\mu_B$ respectively. Here we find that Ti and
V doped Na clusters posses even larger moments. In case of V our results agree
with the experimental study of Bergmann {\it et al.}~\cite{15,16} that predict a high
magnetic moment of 6-7 $\mu_B$ for V as an impurity in bulk Na. These results, we
believe, can provide an important way to produce magnetic clusters with large
moments. In order to convince ourselves that the large magnetic moments we
found on these clusters were not artifacts of the methods we used, and to
further benchmark our calculations, we also calculated the properties of a
CrAu$_6$ cluster in a planar geometry using plane-wave USPP method for which
previous studies exist. In agreement with Li {\it et al.}~\cite{13}, we obtained a moment of
4 $\mu_B$. The molecular orbitals (MO) diagram we found was in very good agreement
with that presented in Fig. 3 of ref. 13.

One of the important questions is the nature of the polarization at the alkali
metal sites. When a transition metal atom is embedded in a bulk free electron
host, it is known that a many body state where the host polarization is
opposed to the transition metal moment can be formed. This leads to the well
known Kondo resonance~\cite{34}. It is then interesting to ask how such an effect is
modified in reduced sizes. In Figure~\ref{fig:1} we give the spin moment on the atoms
obtained by carrying out a Mulliken population analysis of the atomic spin
charges. Note that the host polarization changes in magnitude and sign as the
cluster size is changed. These modifications can be linked to the location of
the electronic states and their interactions between the host and the
transition metal that we discuss below. 

We now try to understand the origin of such larger-than-atom magnetic moments
in these clusters and the nature of the host polarization. Figure~\ref{fig:2} shows the
MO diagram for Na$_4$, TMNa$_4$ and TM atoms. The angular momentum characters of the
MO's are also presented. Not surprisingly, more $d$ type MO's get occupied as
the TM atom is changed from Sc to V. For ScNa$_4$ in the majority $\alpha$ spin channel,
we found that the highest occupied molecular orbital (HOMO), and HOMO-1 have
largely $d$ character on the Sc atom, along with some $s$ and/or $p$ contributions
from the Na atoms (s,p$_{Na}$ d$_{Sc}$ in our nomenclature). In the minority $\beta$ spin
channel, the HOMO level is a mix of $s$ and $p$ type orbitals on Na (s,p$_{Na}$), and $p$
type orbitals on Sc (p$_{Sc}$). The lowest unoccupied MO's (LUMO's) in the $\alpha$ and $\beta$
spin channels are of s,p$_{Na}$ d$_{Sc}$ type. In TiNa$_4$ we found that the HOMO and
HOMO-2 levels presented a s,p$_{Na}$ d$_{Ti}$ character and that HOMO-3 is mainly of d$_{Ti}$
character. VNa$_4$ is the first cluster to exhibit an enhanced magnetic
moment. An analysis of the MO characters shows that the HOMO and 
HOMO-2 levels presented a s,p$_{Na}$ d$_V$ character. HOMO-3 and HOMO-5 are of d$_V$
character. We can observe that in the ScNa$_4$, TiNa$_4$ and VNa$_4$ series the $spd$
hybridization increases, as there is an increasing number of $d$ levels available
at lower energies in the Sc, Ti and V series.

Figure~\ref{fig:3} shows the MO diagram for Na$_5$, TMNa$_5$ and TM atoms. In this series the
TiNa$_5$, VNa$_5$ clusters presented enhanced magnetic moments. An analysis of the
character of the MO's shows a $spd$ hybridization (s,p$_{Na}$ d$_{Sc}$) in the case of
ScNa$_5$. Interestingly, the MO's of TiNa$_5$ and VNa$_5$ present a s,p$_{Na}$ p$_{TM}$ (and not
s,p$_{Na}$ d$_{TM}$) character. In TiNa$_5$ we found that the doubly degenerate HOMO, and
the HOMO-1 levels are mainly of $d$ character. This result shows that in
addition to hybridization there is a charge transfer from the Na $s$ levels to
the Ti $d$ states, as an isolated Ti atom possesses only two filled $d$
states. The same picture was found in VNa$_5$ where the degenerated HOMO and
HOMO-2 levels are of $d$ character. Figure~\ref{fig:4} shows the MO diagram for Na$_6$, TMNa$_6$
and TM atoms. In this series also the clusters with Ti and V presented
enhanced magnetic moments. Analysis of the MO's characters suggests an $spd$
hybridization to be responsible for this enhancement.

We conjecture that the major reason for the enhanced magnetic moments on these
clusters is this significant $spd$ hybridization. In fact, total occupancy of $d$
and $spd$ states in some of these clusters is larger than the occupancy of $d$
states in the respective isolated TM atoms. Normally hybridization of atomic
orbitals leads to quenching of atomic moments. Here, hybridization along with
a large exchange splitting has an opposite effect.  For example, exchange
splitting (defined as the energy difference between the highest $\alpha$-spin MO and
the lowest $\beta$-spin MO having predominantly $d$ characters) in TiNa$_6$ and VNa$_6$
clusters is $\sim 1$ and $\sim 1.8$ eV respectively. 

A comparison of the TM doped Na$_6$ clusters with the TM doped Au$_6$ clusters
reported by Li {\it et al.}~\cite{13} is of some importance here. In the Au$_6$ cluster, the
TM's act like guest atoms. The spectral features of the TMAu$_6$ clusters showed
two distinct regions which could be attributed to a TM atom and to an Au$_6$
ring. This conclusion was borne
out by their DFT calculations which revealed that the MO's near the gap were
mostly of pure $d$-type on the TM atom, and there was a substantial gap between
these levels and the deeper lying levels derived from the Au$_6$ hexagon. In the
present study there is no clear guest-host separation of MO's in TMNa$_n$
clusters. On the contrary, in TiNa$_n$ and VNa$_n$ clusters, at least part of the
moment comes from states that have significant $spd$ mixing as we have
argued. 

Relative values of electronegativity of Na and TM atoms may also have a
crucial role to play. Na has a smaller (Pauling) electronegativity compared to
the TM atoms. Thus, nominally one expects charge transfer from the Na to the
TM atoms. This explains the charge transfer found in the TMNa$_5$ clusters. In
contrast to Na, Au has a (Pauling) electronegativity value much larger than
those of Ti, V and Cr, suggesting that there should be a charge transfer from
TM atoms to Au. Indeed, Torres {\it et al.} ~\cite{14} found in their DFT calculations that
the charge transfer from TM to Au atoms in TMAu$_n^+$ clusters roughly follows the
sequence of electronegativity difference between Au and TM atoms, and for
small cluster sizes the Au-TM bond is mainly ionic. The amount of charge
transfer from the TM to the Au atoms determines the moment on the TM atom. In
TMNa$_n$ clusters, as we have already shown, overlap of Na $s$ and $p$ and TM $d$ type
orbitals increases the moment on the TM atoms.
 
The relative stability of a TMNa$_n$ clusters can be measured by the embedding
energy (EE) of the TM in the Na$_n$ cluster. This is defined as, 
\begin{equation}
  \label{eq:ec}
  EE({\rm TMNa}_n) = E({\rm Na}_n) + E({\rm TM}) - E({\rm TMNa}_n), 
\end{equation}
\noindent
where E(TMNa$_n$) is the ground state total energy of a TM doped Na$_n$ cluster,
E(Na$_n$) is the total energy of a pure Na$_n$ cluster, and E(TM) is the total
energy of an isolated TM atom. It has been shown before that it is important
to preserve spin multiplicity while calculating embedding energies~\cite{35,36}.
Therefore, in calculating EE we used total energies of excited states for TM
and Na$_n$ when the spin multiplicities of the ground states did not satisfy the
Wigner-Witmer spin conservation rule~\cite{36}. In calculating the EE for ScNa$_n$,
TiNa$_4$, TiNa$_5$, and VNa$_5$ the ground state energies were employed. In the rest of
the cases, excited states of either Na$_n$ cluster or the TM atom has been
considered, whichever gives a larger EE. The calculated EE for the TMNa$_n$
series are presented in Table~\ref{Table:1}.

It is interesting to note certain trends in this family of clusters. For each
size from Na$_4$ to Na$_6$, EE increases as the TM atom is changed from Sc to Ti and
V. For any given TM atom, the EE increases from Na$_4$ to Na$_5$ and Na$_6$. The second
fact suggests that in experiments, these small clusters will tend to bind with
other Na atoms and form larger clusters. Therefore, a study extending up to
larger TMNa$_n$ clusters is of importance, and that is currently underway. A
crucial question is, up to what size of Na clusters such large moments survive
or how the moment evolves for a given TM atom, as the number of Na atoms
increases. This study will allow us to find out if Na clusters are able to
present large magnetic moments in the range of 6.6 to 7 $\mu_B$ as found in the
experimental studies of Bergmann {\it et al.}~\cite{15,16}. Our preliminary calculations show
that large moments survive on the Ti and V doped Na$_n$ clusters at least up to $n
= 9$~\cite{37}. 

In order for any cluster to be technologically useful, it has to be chemically
inert. One indication of chemical inertness is the gap between the HOMO and LUMO
levels. HOMO-LUMO gaps for these clusters are shown in Table~\ref{Table:1} . All the
clusters have a large gap of $\sim 0.5$ eV or more. Some of them, in fact, have
gaps in excess of 1 eV. The large EE values and HOMO-LUMO gaps are indicative
of stability in these clusters, and reinforce the idea that they may have
useful applications.

\section{Conclusion}
The present studies show that the size and structure of alkali metal clusters
can significantly affect the spin magnetic moments of doped transition metal
atoms. In particular, V doped Na$_4$, and Ti and V doped Na$_5$ and Na$_6$ clusters are
shown to exhibit enhanced magnetic moments. The present results for VNa$_n$
clusters support the conclusions of the experimental studies by Bergman {\it et
al.} ~\cite{15,16} that the V atoms supported on alkali layers can attain large spin
moments. ScNa$_6$ cluster also has a large moment of 
3 $\mu_B$ in a low lying isoenergetic state. An analysis of the MO's of TMNa$_n$
clusters suggests that a significant hybridization between Na $s$ and $p$ and TM $d$
type orbitals coupled with a large exchange splitting of $d$ type states, and a
smaller value of electronegativity of Na compared to those of the TM atoms are
responsible for the observed large moments and the host polarization. This
conjecture is consistent with earlier calculations on neutral and cationic
TMAu$_6$ clusters. We would like to add that the host polarization can be
expected to stabilize the orientation of the transition metal moment providing
spin anisotropy with respect to the geometry of the alkali host. These effects
are being investigated.

\section{Acknowledgments}
J.\ U.\ R. acknowledges support from U.\ S. Air Force Office of Scientific Research
grant FA9550-05-1-0186, while S.\ N.\ K. is grateful to U.S. Department of Energy
Grant DE-FG02-96ER45579 for support. VASP numerical calculations for this
study were carried out at the cluster computing facility in the Harish-Chandra
Research Institute (http://cluster.mri.ernet.in).

\vspace*{0.5in}


\begin{table}[h]
\caption{ Electronic properties of the TMNa$_n$ clusters. Magnetic moments
  ($\mu$), HOMO-LUMO gaps (Gap), and the embedding energies (EE) in the ground
  state. Magnetic moment in the low-lying isomer, and the energy difference
  with respect to the ground state ($\Delta E$).} 
\begin{tabular}{|c|cccc|cc|}
\hline
Cluster  &  & Ground  state   & &   & Low-lying    & isomer   \\  
  &  $\mu$($\mu_B$) & Gap (eV) & & EE (eV) & $\mu$($\mu_B$) & $\Delta E$ (eV)  \\ 
\hline
ScNa$_4$ &  1 &  0.65 & &  1.10 &  1 & 0.025  \\
TiNa$_4$ &  2 &  1.03 & &  1.17 &  4 & 0.013  \\
VNa$_4$  &  5 &  0.41 & &  1.73 &  3 & 0.064  \\
ScNa$_5$ &  0 &  0.51 & &  1.32 &  2 & 0.003  \\
TiNa$_5$ &  3 &  0.92 & &  1.50 &  3 & 0.104  \\
VNa$_5$  &  4 &  1.17 & &  1.89 &  4 & 0.283  \\
ScNa$_6$ &  1 &  0.67 & &  1.56 &  1 & 0.076  \\
TiNa$_6$ &  4 &  0.48 & &  1.89 &  2 & 0.030  \\
VNa$_6$  &  5 &  0.68 & &  2.21 &  3 & 0.128  \\
\hline
\end{tabular}
\label{Table:1}
\end{table}

\vspace*{0.5in}
\centerline{\bf Figure Captions}

 \begin{figure}[h]
 \caption{ (Color online) Ground state geometries (GS), and immediate higher
 energy isomers (EX) of the TMNa$_n$ clusters. The bond lengths are given in Angstroms
 and superscripts indicate spin multiplicity. The Mulliken atomic spin charges
 for the symmetry inequivalent atoms are marked below them.}
 \label{fig:1}
 \end{figure}

 \begin{figure}[h]
 \caption{One-electron levels in Na$_4$, ScNa$_4$, Sc, TiNa$_4$, Ti, VNa$_4$ and V,
  superscripts indicate the spin multiplicity. The continuous lines represent
  occupied levels; the dotted lines correspond to unfilled states. The
  degeneracy is marked by a number next to the level. The angular characters
  of the energy levels are also given. The arrows indicate the majority (up),
  and minority (down) spin states.}
 \label{fig:2}
 \end{figure}

 \begin{figure}[h]
 \caption{One-electron levels in Na$_5$, ScNa$_5$, Sc, TiNa$_5$, Ti, VNa$_5$ and V,
 superscripts indicate the spin multiplicity. The continuous lines represent
 occupied levels; the dotted lines correspond to unfilled states. The
 degeneracy is marked by a number next to the level. The angular character of
 the energy levels is also given.  The arrows indicate the majority (up), and
 minority (down) spin states. }
 \label{fig:3}
 \end{figure}

 \begin{figure}[h]
 \caption{One-electron levels in Na$_6$, ScNa$_6$, Sc, TiNa$_6$, Ti, VNa$_6$ and V
 superscripts indicate the spin multiplicity. The continuous lines represent
 occupied levels; the dotted lines correspond to unfilled states. The
 degeneracy is marked by a number next to the level. The angular character of
 the energy levels is also given. The arrows indicate the majority (up), and
 minority (down) spin states.}
 \label{fig:4}
 \end{figure}

\end{document}